\documentclass[a4paper,11pt]{article}
\usepackage{pos}
\usepackage{subcaption}
\usepackage{CJKutf8}

\begin{document}

\title{Helicity Dependent Distribution Functions of the Proton and $\Lambda$ and $\Sigma^0$ Baryons}


\author[a]{Yang Yu%
       $\,^{\href{https://orcid.org/0009-0008-8011-3430}{\textcolor[rgb]{0.00,1.00,0.00}{\sf ID}},}$}

\author[b]{Peng Cheng%
    $\,^{\href{https://orcid.org/0000-0002-6410-9465}{\textcolor[rgb]{0.00,1.00,0.00}{\sf ID}},}$}

\author[c,d]{Hui-Yu Xing%
    $\,^{\href{https://orcid.org/0000-0002-0719-7526}{\textcolor[rgb]{0.00,1.00,0.00}{\sf ID}},}$}

\author[e]{\\Daniele Binosi%
    $\,^{\href{https://orcid.org/0000-0003-1742-4689}{\textcolor[rgb]{0.00,1.00,0.00}{\sf ID}},}$}
    
\author*[c,d]{Craig D.\ Roberts%
       $^{\href{https://orcid.org/0000-0002-2937-1361}{\textcolor[rgb]{0.00,1.00,0.00}{\sf ID}},}$}

\affiliation[a]{School of Science, \href{https://ror.org/05x2f1m38}{East China Jiaotong University}, Nanchang, Jiangxi 330013, China}

\affiliation[b]{Department of Physics, \href{https://ror.org/05fsfvw79}{Anhui Normal University}, Wuhu, Anhui 24100, China}

\affiliation[c]{School of Physics, \href{https://ror.org/01rxvg760}{Nanjing University},
Nanjing, Jiangsu 210093, China}

\affiliation[d]{Institute for Nonperturbative Physics, \href{https://ror.org/01rxvg760}{Nanjing University}, Nanjing, Jiangsu 210093, China}

\affiliation[e]{European Centre for Theoretical Studies in Nuclear Physics
            and Related Areas  (\href{https://ror.org/01gzye136}{ECT*})\\ \hspace*{0.5em}Villa Tambosi, Strada delle Tabarelle 286, I-38123 Villazzano (TN), Italy}

\emailAdd{cdroberts@nju.edu.cn}




\abstract{
\vspace*{-10ex}
\rightline{\sf NJU-INP 113/26}

\vspace*{9ex}
Using continuum Schwinger function methods, a coherent set of predictions for proton, $\Lambda$ and $\Sigma^0$ distribution functions (DFs) has been made available -- both helicity dependent and unpolarised.  The results and comparisons between them reveal impacts of diquark correlations and SU$(3)$-flavour symmetry breaking, some of which are highlighted in this contribution.  For instance: in-proton ratios of helicity-dependent/unpolarised valence-quark DFs are presented; it is highlighted that, were it not for the presence of axialvector diquarks in the $\Sigma^0$, the valence strange quark would carry none of the $\Sigma^0$ spin; and the sign and size of polarised gluon DFs is discussed -- at a scale typical of modern measurements, gluon partons carry roughly 40\% of each octet baryon's spin.
}

\FullConference{The 26th International Symposium on Spin Physics (SPIN25) -- A Century of Spin\\
2025 September 22-26\\
Qingdao Haitian Hotel, Qingdao, Shandong Province, China\\}


\maketitle

\section{Foundation}
High-energy nuclear and particle physics communities worldwide are working to establish whether quantum chromodynamics (QCD) is the theory that explains strong interactions in the Standard Model (SM).  As part of that effort, they have challenged themselves to answer the following three of Nature's most fundamental questions:
(\emph{a}) What is the origin of the mass-scale, $m_N$, the proton mass, which sets the basic unit of mass for all visible matter and, whatever it is, why is the pion, with its unnaturally small, lepton-like mass, $m_\pi \approx m_N/7 \approx m_{\rm muon}$, seemingly oblivious;
(\emph{b}) Given that the proton is composite, how is its measurable $J=1/2$ spin distributed amongst its constituents;
and (\emph{c}) Accepting that self-interactions amongst gluon partons are the basic expression of QCD's non-Abelian character, what impacts do they have on both gluons themselves and observable quantities?
It is anticipated that data obtained from a diverse array of modern and foreseen facilities will pave the way to solutions \cite{Denisov:2018unjF, Aguilar:2019teb, Anderle:2021wcy, Arrington:2021biu, Quintans:2022utc, Accardi:2023chb, Achenbach:2025kfx, Lu:2025bnm, Messchendorp:2025men}.
The answers are important because the characterising mass scale emerged roughly $1\,\mu$s after the Big Bang and thereafter had a critical influence on the evolution of the Universe.

The Lagrangian defining QCD can be traced back to Ref.\,\cite{Fritzsch:1973pi}, in which it was stated:
``\emph{The quarks come in three `colors,' but all physical states and interactions are supposed to be singlets with respect to the SU$(3)$ of color. Thus, we do not accept theories in which quarks are real, observable particles; nor do we allow any scheme in which the color non-singlet degrees of freedom can be excited}.''  So, from the beginning, strong-interaction practitioners were confronted by a battery of basic challenges.
If gluons and quarks are not observable, then what are the (asymptotic) detectable degrees-of-freedom;
and how are they built from the Lagrangian degrees-of-freedom?
Moreover, if the degrees-of-freedom used to write the QCD Lagrangian are not directly measurable, then something essentially nonperturbative underlies all observable strong interaction physics.  That being the case, what theoretical tools can be used to test if QCD is really the theory of strong interactions?  Finally and more fundamentally still, given that QCD is a Poincar\'e-invariant quantum gauge field theory in four dimensions, is it really a theory or just one more effective field theory?  The answer to this last question could have implications that reach far beyond the SM.

The process of connecting existing and expected data with QCD predictions should be objective, \emph{i.e}.,  independent of reference frame, measurement process, and other subjective (observer-dependent) considerations \cite{Brodsky:2022fqy}.
Owing to the deeply nonperturbative character of QCD, this requirement places great demands on phenomenology and theory, neither of which is yet fully able to meet them.
Consequently, today, there is much debate over the interpretations of even existing data and their merits as tests of QCD.

\section{Emergent Hadron Mass}
When insisting on an objective approach, then the proton mass, $m_N$, can be decomposed into three distinct component contributions.
[\emph{i}] Higgs boson only.  This part is the amount of $m_N$ that can unambiguously be attributed to Higgs-boson generated renormalisation group invariant (RGI) quark current masses: $\hat m_H = 2\hat m_u + \hat m_d \approx 13\,$MeV\,$=0.014\,m_N$.  Plainly, this falls far short of the target value.
(The concept of a RGI current mass is as old as QCD \cite{Politzer:1976tv}.)
[\emph{ii}] Constructive interference with non-Higgs mass generating mechanisms.  Using QCD-connected theory or experimental data, one can determine the magnitude of the in-nucleon expectation value of the RGI mass term in the QCD Lagrangian \cite{Flambaum:2005kc, Alarcon:2021dlz}:
\begin{equation}
\hat \sigma_N := \langle N(k) | \int d^4 x\, \hat m [\widehat{\bar q(x) q(x)}] |N(k)\rangle \approx 60\,{\rm MeV}\,.
\end{equation}
Thus, constructive interference between Higgs-alone and other mass generating mechanisms yields $\hat m_\sigma = \hat \sigma_N - \hat m_H  
\approx (60-13)\,{\rm MeV} = 0.05\,m_N$.  Again, one is left far from the desired value.
The remainder is [\emph{iii}] emergent hadron mass (EHM), \emph{viz}.\ $\hat m_{\rm EHM} \approx 0.94\,m_N$.
Evidently, the challenge posed in (\emph{a}) above is either to uncover the QCD origin of
\begin{equation}
\hat m_\sigma + \hat m_{\rm EHM} \approx 0.99 m_N\,,
\end{equation}
or give up on QCD as the underlying theory of strong interactions.
(N.B.\ Since the nucleon contains no valence $s$ quarks, then any $s$-quark $\hat\sigma_N$-like term is properly included in $\hat m_{\rm EHM}$.)

An explanation of EHM within QCD was proposed in Ref.\,\cite{Cornwall:1981zr}.  Stated simply: owing to strong self-interactions, enabled by QCD's $3$- and $4$-gluon vertices, gluon partons become gluon quasiparticles whose propagation is characterised by a RGI momentum-dependent mass function that is large at infrared momenta.  The associated RGI gluon mass scale is $\hat m = 0.43(1)\,$GeV \cite{Binosi:2016nme, Cui:2019dwv}, \emph{viz}.\ roughly one-half the proton mass.  This phenomenon lies at the heart of the contemporary EHM paradigm \cite{Roberts:2021nhw, Binosi:2022djx, Ding:2022ows, Roberts:2022rxm, Ferreira:2023fva, Carman:2023zke, Raya:2024ejx}.

One critical corollary of the gluon mass is the emergence in QCD of a unique-analogue of the Gell-Mann--Low effective charge in quantum electrodynamics (QED) \cite{GellMann:1954fq}.
This is the archetypal running coupling that is now known with a precision that places the uncertainty in the twelfth significant figure \cite{ParticleDataGroup:2024cfk}.
Like the QED running coupling, the QCD effective charge \cite{Binosi:2016nme, Cui:2019dwv, Brodsky:2024zev}, $\hat\alpha(k^2)$, is process-independent.
Furthermore, it has many other properties, \emph{inter alia}, exhibiting conformal behaviour at infrared momenta with $\hat\alpha(k^2=0)/\pi=0.97(4)$ -- see, \emph{e.g}., Ref.\,\cite[Fig.\,3]{Ding:2022ows} and associated discussion, which make it an ideal candidate for that long-sought running coupling which defines the interaction strength in QCD at all length scales \cite{Dokshitzer:1998nz}.

The running gluon mass and process-independent effective charge combine to form a key component of the kernel in the quark gap (Dyson) equation.  The remaining piece is defined by the gluon + quark vertex \cite{Binosi:2016wcx}.  The gap equation solution obtained therewith is the third pillar of EHM, \emph{i.e}., the running quark mass, $M(k^2)$; see, \emph{e.g}., Ref.\,\cite[Fig.\,2.5]{Roberts:2021nhw}.
Thus the quark partons of QCD's Lagrangian are also transmogrified, becoming dressed-quark quasiparticles characterised, too, by a RGI mass function. 
In the light quark sector, one finds $M(0) \approx m_N/3$; and, therewith, via the Faddeev equation approach to the nucleon bound-state problem \cite{Eichmann:2009qa, Yao:2024ixu}, one immediately sees the origin of the nucleon mass scale in QCD.

\section{Distribution Functions - Proton}
Continuum Schwinger function methods (CSMs) have been employed to deliver, amongst other things, a tenable species separation of nucleon gravitational form factors \cite{Yao:2024ixu},
a large array of unified predictions for all pion, kaon, and proton distribution functions (DFs) \cite{Cui:2020tdf, Chang:2022jri, Lu:2022cjx, Cheng:2023kmt, Yu:2025fer} and
for pion and kaon fragmentation functions (FFs) \cite{Xing:2025eip}, 
and useful information on quark and gluon angular momentum contributions to the proton spin \cite{Yu:2024ovn}.  
Of particular interest, perhaps, are the predictions for 
$\bar d/\bar u$ and $F_2^n/F_2^p$ and the associated comparisons with data \cite[Fig.\,4]{Lu:2022cjx};
and predictions for the gluon contribution to the proton spin \cite{Cheng:2023kmt, Yu:2024ovn}, which reveal that, in the infinite-momentum frame, 40\% of the proton spin is carried by gluon partons at a scale typical of modern measurements.

\begin{figure}[t]
    \centering
    \begin{subfigure}{0.475\columnwidth}
        \includegraphics[width=\textwidth]{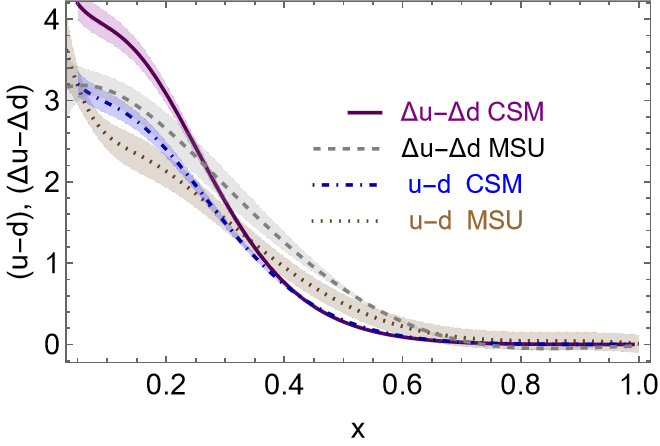}
        \caption{\small Isovector DFs}
        \label{fig3:subim1}
    \end{subfigure}
    \begin{subfigure}{0.49\columnwidth}
        \includegraphics[width=\textwidth]{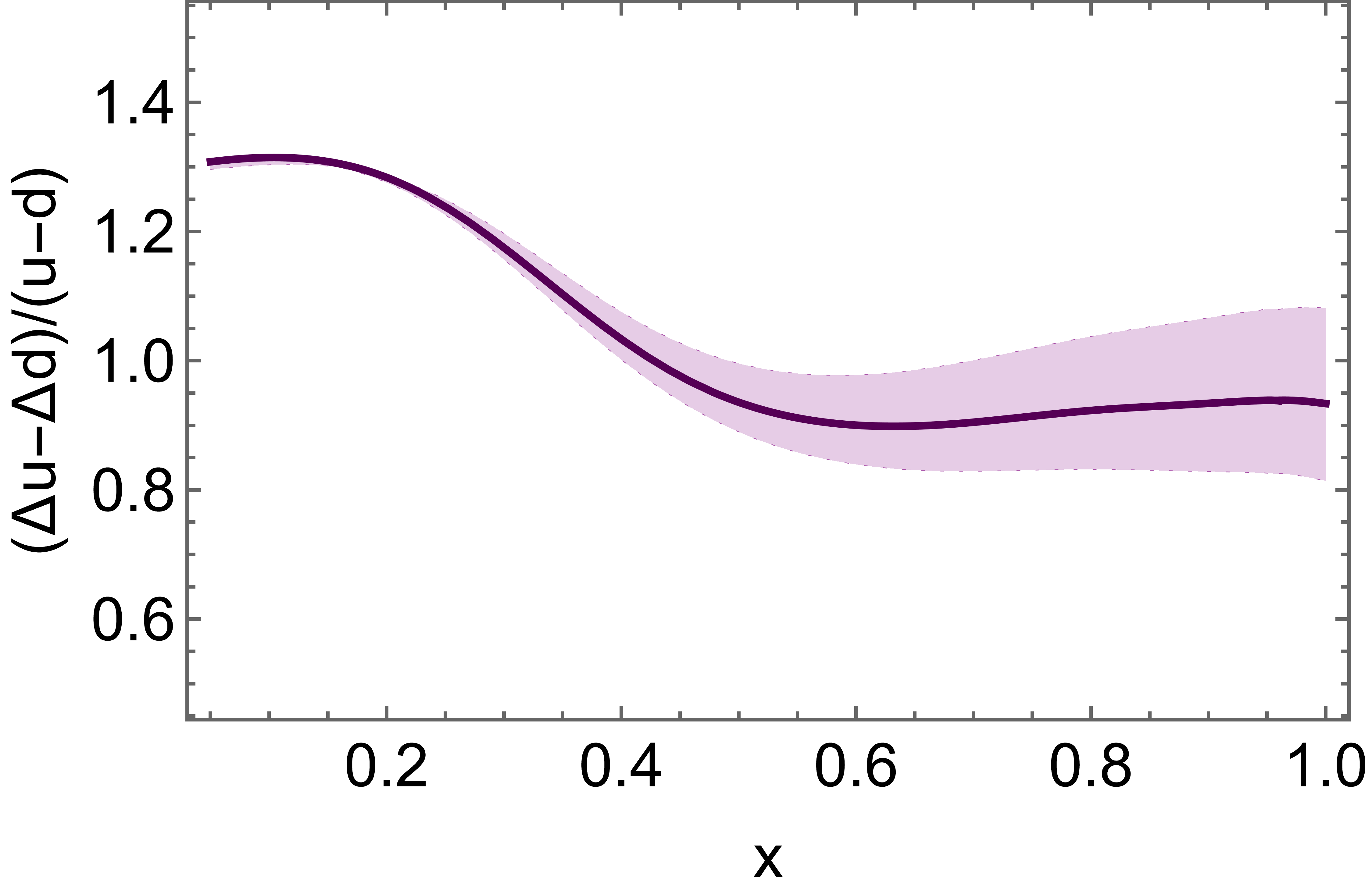}
        \caption{\small Isovector Ratio}
        \label{fig3:subim2}
    \end{subfigure}
    \vspace*{1ex}
    
    \begin{subfigure}{0.49\columnwidth}
        \includegraphics[width=\textwidth]{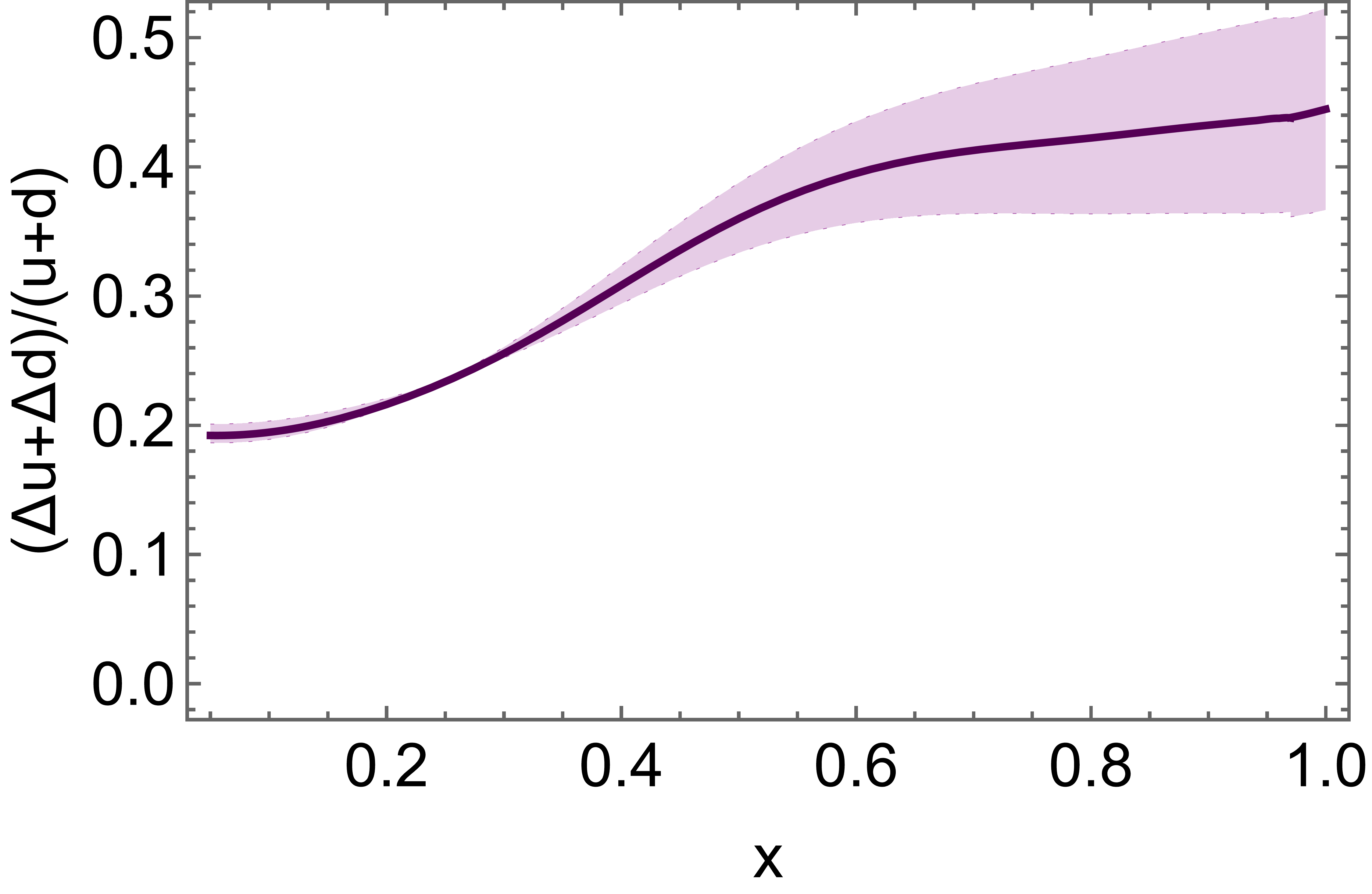}
        \caption{\small Isoscalar Ratio}
        \label{fig3:subim3}
    \end{subfigure}
    \caption{CSM predictions for a collection of in-proton DFs, all CSM results at the scale $\zeta=2\,$GeV.
    {\sf Panel a}.  Isovector in-proton DFs compared with preliminary results from numerical simulations of lattice-regularised QCD \cite{Lin:2020fsj, Lin:2018pvv}.
    {\sf Panels b, c}.  CSM predictions for in-proton isovector and isoscalar DF ratios.
\label{F1T}}
\end{figure}

Some CSM predictions for in-proton unpolarised and polarised DFs are depicted in Fig.\,\ref{F1T}.  The CSM results were calculated using the framework in Ref.\,\cite{Cheng:2023kmt}.  Notably, empirical information on the ratios in Panels b, c may soon be forthcoming from Jefferson Laboratory.

\section{Distribution Functions - $\mathbf \Lambda$ and $\mathbf \Sigma^{\mathbf 0}$}
Comparisons between structural properties of $\Lambda$ and $\Sigma^0$ baryons is of special interest because they have the same valence-quark content but different isospin: $\Lambda (I=0)$ and $\Sigma^0(I=1)$.
Consequently, in the modern quark + diquark picture of baryons \cite{Barabanov:2020jvn, Yu:2025fer, Cheng:2025yij}, their spin-flavour wave functions have markedly different structure:
\begin{equation}
\label{FaddeevFlavour}
u_\Lambda  =
\frac{1}{\sqrt 2}
\left[
\begin{array}{ll}
{\rm r}_1^0\,: & \sqrt{2}s[ud]_{0^+}^{I=0}       \\
{\rm r}_2^0\,: & d[us]_{0^+}^{I=0} - u[ds]_{0^+}^{I=0}   \\
{\rm r}_3^0\,: & d\{us\}_{1^+}^{ I=1} - u\{ds\}_{1^+}^{ I=1}
\end{array}
\right] , \;\;
u_{\Sigma^0}  =
\frac{1}{\sqrt 2}
\left[
\begin{array}{llr}
{\rm r}_1^1\,: & d [us]_{0^+}^{I=0} + u [ds]_{0^+}^{I=0}       \\
{\rm r}_2^1\,: & 2 s\{ud\}_{1^+}^{I=1}     \\
{\rm r}_3^1\,: & d\{us\}_{1^+}^{I=1} + u\{ds\}_{1^+}^{I=1}
\end{array}
\right] . 
\end{equation}
These column-vectors show that the $\Lambda$ has two distinct isoscalar-scalar diquark components ($I=0, J^P=0^+$) and one isovector-axialvector, whereas the $\Sigma^0$ has one isoscalar-scalar contribution and two isovector-axialvector.
(Scalar and axialvector refers to the mapping properties of the diquarks under Lorentz transformations.)
It is likely that these differences are expressed in the unpolarised and helicity dependent DFs of these systems.

With this motivation, parton DFs for the $\Lambda$, $\Sigma^0$ baryons were calculated in Ref.\,\cite{Yu:2025fer}  -- valence, glue, and four-flavour separated sea, unpolarised and polarised -- and therein compared with those of nucleons.  The study used the quark + dynamical diquark picture of baryons realised via a symmetry-preserving treatment of a vector $\otimes$ vector contact interaction (SCI).  Widespread use indicates that SCI predictions provide a fair guide to those that would be obtained from analogous approaches to baryon structure that use Schwinger functions which more realistically express fundamental features of QCD, \emph{i.e}., those determined by a momentum-dependent interaction.
A strength of the SCI is that typical analyses are largely algebraic; hence, the formulae and results are readily understood.  This enables judgements to readily be made concerning both SCI outcomes themselves and, via relevant comparisons, results obtained using more sophisticated frameworks.

Considering the far-valence domain, Ref.\,\cite{Yu:2025fer} found that it is only with the presence of axialvector diquarks in the $\Sigma^0$ that one obtains finite values for $l(x)/s(x)|_{x\simeq 1}$, $\Delta l(x)/\Delta s(x)|_{x\simeq 1}$, $l = u, d$.
Even with axialvector diquarks, the dominance of scalar diquarks in the $\Sigma^0$ leads to values of these ratios that are large compared with their $\Lambda, p$ analogues.
Regarding the $\Sigma^0$ spin-flavour amplitude, Eq.\,\eqref{FaddeevFlavour}\,-\,right, only with the existence of axialvector diquarks can the $s$ quark participate as a valence degree of freedom in the $\Sigma^0$; otherwise, it is always ``hidden'' in an isoscalar-scalar diquark correlation.
Hence, in the absence of axialvector diquarks, the valence $s$ quark contribution to the $\Sigma^0$ spin is zero.

Higgs couplings into QCD generate a very large disparity between $l$- and $s$-quark current masses.
However, Ref.\,\cite{Yu:2025fer} revealed that in baryons, as in mesons, this Higgs-driven imbalance is largely masked by the size of EHM effects.
Thus, whilst the peak locations in $x$-weighted $\Lambda, \Sigma^0$ valence quark DFs are shifted with respect to those in the nucleon, the relocations are modest and reflect the impacts of both diquark structure and dressed-quark mass differences.

Naturally, glue and sea DFs are nonzero.
Regarding helicity-independent glue DFs, Ref.\,\cite{Yu:2025fer} found that the profiles in $\Lambda$, $\Sigma^0$, $p$ are similar.
Differences only become apparent when one considers ratios, which reveal that, on the valence quark domain, both the glue-in-$\Lambda$ and glue-in-$\Sigma^0$ DFs are suppressed compared to the glue-in-$p$ DF: the in-$\Sigma^0$ suppression is greater.
Helicity independent sea DFs are semiquantitatively similar to glue DFs.
The SCI analysis predicts a nonzero $c+\bar c$ DF in each hadron, \emph{viz}.\ $\Lambda, \Sigma^0, p$: in each case, they have sea quark profiles and roughly commensurate magnitudes.
Thus one does not need to invoke ``intrinsic charm'' \cite{Brodsky:1980pb} in order to explain existing phenomenological inferences of the strength of in-proton $c+ \bar c$.

Helicity dependent glue and sea DFs are also nonzero.
In each hadron, the helicity-dependent glue DF is non-negative; and compared with kindred unpolarised glue DFs, the peak magnitudes of the $x$-weighted polarised DFs are approximately an order of magnitude smaller.
The SCI predicts that the $\Lambda$ contains more polarised glue than the proton.  
There is less in the $\Sigma^0$.
However, these differences are small.
Helicity-dependent $x$-weighted sea DFs are an order-of-magnitude smaller than their unpolarised partners.

With all DFs available, it was possible to discuss $\Lambda, \Sigma^0, p$ spin decompositions.
Incorporating the non-Abelian anomaly contribution to the flavour-singlet current, the SCI provides a viable explanation of proton spin measurements, with predicted values for each hadron lying within extant empirical bounds.
In detail, at $\zeta_2$, compared with the proton value, quarks carry 12\% more of the $\Lambda$ spin and $3$\% less of that of the $\Sigma^0$.
The net glue contribution to each baryon's spin is the same, \emph{viz}.\ $\approx 40$\%;
and regarding quark orbital angular momentum, that in the $\Lambda$-baryon is only $\approx 55$\% of that in $\Sigma^0, p$; see, also, Ref.\,\cite{Cheng:2023kmt}.

The SCI study in Ref.\,\cite{Yu:2025fer} completed an important step in a systematic programme that aims to deliver QCD-connected predictions for all baryon DFs, their unification with those of mesons, and the complete set of related FFs.
It would also be worth developing the framework further so that one could deliver predictions for spin transfer in deep-inelastic $\Lambda$ electroproduction.  Data exist \cite{HERMES:1999buc, Schnell:2024ppe} and more are expected from modern and anticipated facilities.  However, the absence of reliable information on the fragmentation functions involved is an impediment to such an extension \cite{Metz:2016swz}; hence, the goal listed above.
Analyses of DFs characterising baryons containing one or more heavy quarks may also be valuable. 
They would, for instance, expose the impact on EHM expressions made by stronger Higgs boson couplings into QCD.

\section{Conclusion}
The past decade has seen real progress in strong interaction theory.  
One now sees an expanding array of parameter-free predictions for the proton and, significantly, for many of the other hadrons whose properties express the full meaning of QCD.  
We would like to stress, \emph{e.g}.,  that insights are being drawn into the structure of Nature's most fundamental Nambu-Goldstone bosons -- pions and kaons -- and also the structure of nucleon resonances \cite{Cheng:2025sdp}.

A powerful motivation for all these efforts is the need to understand how the apparently simple QCD Lagrangian can explain the confinement of colour and the emergence of the diverse and complex array of detectable hadronic states.  One may expect that the precise data needed to test any conjecture will become available during operations of existing and anticipated high-luminosity, high-energy facilities.
A validated explanation will move science into a new realm of understanding by, possibly, proving QCD to be the first well-defined four-dimensional quantum field theory ever proposed.  
If such is the case, then QCD's EHM paradigm may open doors that lead far beyond the Standard Model.

\section*{Acknowledgments}
We thank X.-C.\ Zheng for useful suggestions.
Work supported by:
National Natural Science Foundation of China (grant no.\ 12135007);
and
Natural Science Foundation of Anhui Pro\-vince, grant no.\ 2408085QA028.


\end{document}